\documentclass[11pt]{article}

\usepackage[english]{babel}

\usepackage[a4paper,top=2cm,bottom=2cm,left=2.5cm,right=2.5cm,marginparwidth=1.75cm]{geometry}
\usepackage{setspace}
\usepackage{graphicx}
\usepackage[colorlinks=true, allcolors=blue]{hyperref}
\usepackage[utf8]{inputenc}
\usepackage[T1]{fontenc}
\usepackage[english]{babel}
\usepackage{amsmath,amssymb}
\usepackage[numbers]{natbib}
\usepackage{hyperref}
\usepackage{titlesec}
\usepackage{caption}
\usepackage{setspace}
\usepackage{multicol}
\usepackage{ragged2e} 
\usepackage[dvipsnames]{xcolor} 
\usepackage{tcolorbox} 

\newcommand{\hi}{H~{\sc i}} 
\newcommand{\hii}{H~{\sc ii}} 
\newcommand{\hei}{He~{\sc i}} 
\newcommand{\heii}{He~{\sc ii}} 
\newcommand{\cii}{C~{\sc ii}} 
\newcommand{\civ}{C~{\sc iv}} 
\newcommand{\oi}{O~{\sc i}} 
\newcommand{\oii}{O~{\sc ii}} 
\newcommand{\oiii}{O~{\sc iii}} 
\newcommand{\nii}{N~{\sc ii}} 
\newcommand{\sii}{S~{\sc ii}} 
\newcommand{\cliii}{Cl~{\sc iii}} 
\newcommand{\ariv}{Ar~{\sc iv}}
\newcommand{\nev}{Ne~{\sc v}}
\newcommand{\feiii}{Fe~{\sc iii}}
\newcommand{\nel}{$n_{\rm e}$} 
\newcommand{\tel}{$T_{\rm e}$} 
\newcommand{\ts}{$t^2$} 

\title{Why the Northern Hemisphere Needs a 30–40 m Telescope and the Science at Stake: \\Northern Local Star-forming Dwarf Galaxies.  Analogues of the First Galaxies and Probes of the Cosmic Metallicity Scale}
\author{C. Esteban$^{*,1,2}$, J.~M. Vilchez$^{\dagger,3}$, J. Garc\'\i a-Rojas$^{1,2}$, R. Amor\'in$^{3}$,\\
K.~Z. Arellano-C\'ordova$^{4}$, L. Carigi$^{5}$, F. Cullen$^{4}$, O.~V. Egorov$^{6}$\\
S.~R. Flury$^{4}$, J. Iglesias-P\'aramo$^{3}$, C. Kehrig$^{3}$, K. Kreckel$^{6}$\\
J.~E. M\'endez-Delgado$^{5}$, E. P\'erez-Montero$^{3}$, F. F. Rosales-Ortega$^{7}$\\
D. Scholte$^{4}$, T.~M. Stanton$^{4}$, E. Villaver$^{1,2}$\\
[6pt]
{\small \texttt{$^{*}$cel@iac.es}}\\
{\small \texttt{$^{\dagger}$jvm@iaa.es}}\\
[6pt]
{\small $^{1}$Instituto de Astrof\'isica de Canarias, E-38205 La Laguna, Tenerife, Spain}\\
{\small $^{2}$Departamento de Astrof\'isica, Universidad de La Laguna,}\\
{\small E-38206 La Laguna, Tenerife, Spain}\\
{\small $^{3}$Instituto de Astrof\'isica de Andaluc\'ia, IAA-CSIC, Glorieta de la Astronom\'ia S/N,}\\ 
{\small Granada, 18008, Spain}\\
{\small $^{4}$Institute for Astronomy, University of Edinburgh, Royal Observatory,}\\
{\small Edinburgh, EH9 3HJ, United Kingdom}\\
{\small $^{5}$Instituto de Astronom\'{\i}a, Universidad Nacional Aut\'onoma de M\'exico,}\\
{\small Ap. 70-264, 04510 CDMX, Mexico}\\
{\small $^{6}$Astronomisches Rechen-Institut, Zentrum f\"{u}r Astronomie der Universit\"{a}t Heidelberg,}\\
{\small  M\"{o}nchhofstra\ss e 12-14, D-69120 Heidelberg, Germany}\\
{\small $^{7}$Instituto Nacional de Astrof\'isica, \'Optica y Electr\'onica (INAOE),}\\
{\small Luis E. Erro 1, Tonantzintla, 72840 Puebla, Mexico}\\
}

\begin{document}
\maketitle

\newpage

\begin{abstract}

\begin{tcolorbox}[colback=RoyalBlue!5!white,colframe=black!75!black] 
\justifying
{\em  \noindent Star-forming dwarf galaxies in the local Universe, especially extremely metal-poor ones, can be considered analogous to early galaxies of the Epoch of Reionization ($z$ $\geq$ 6). Currently available telescopes cannot adequately detect and measure heavy element recombination lines and certain faint collisionally  excited lines, which are essential for exploring the effects and biases that potential inhomogeneities in electron temperature and density of the ionized gas may have on determining the chemical composition  of these galaxies. On the other hand, the origin of very high-ionization lines (e.g. \heii, [\nev], \civ]) measured in the spectra of an important fraction of these objects remains unknown and a challenge to current stellar models, suggesting the presence of Population III-like stars and/or the existence of non-conventional ionizing sources. 
Obtaining very deep spectra for a selected sample of local star-forming dwarf galaxies would provide unprecedented constraints on their nature, ionization and true chemical abundances, and could change the metallicity scale we assume to understand the chemical evolution of galaxies over cosmic time.}

\end{tcolorbox}

\end{abstract}

\section{Introduction and Motivations}
Understanding the star-formation history of galaxies requires a detailed analysis of how stellar mass assembly occurs across cosmic time. A key tool for studying galaxy evolution is the chemical enrichment of the interstellar medium (ISM) produced by successive generations of stars, whose nucleosynthetic products offer insight into the processes driving the baryon cycle. Local star-forming dwarf galaxies (SFDGs) contain a large mass of ionized gas characterized by an emission-line spectrum. 
The most extreme SFDGs in the Local Universe include the best local analogues to reionization galaxies, such as HII galaxies, blue compact dwarfs (BCDs) or green pea (GP) galaxies \cite{Sargent:70,Terlevich:91,Amorin:12}. They are exceedingly rare compact starbursts, including the most extremely metal-poor galaxies (EMPGs, $<$10\% solar) known \cite{Papaderos:08}. With remarkably similar properties to primeval galaxies (e.g. lowest metallicities, highest star formation per unit mass, and highly ionized and turbulent ISM), they provide excellent local analogues and ideal laboratories for studying massive star formation, chemical enrichment, feedback processes, and the baryon cycle in environments that resemble those frequently observed at higher redshifts \cite{Berg:22,Schaerer:22}. 
Precise measurements of the chemical abundances and physical conditions in such local analogues open a pathway to understanding the evolution of galaxies over cosmic time. 

The chemical composition of SFDGs is usually derived from the intensity of collisionally excited lines (CELs). The abundance of oxygen (O) --the most abundant heavy element in the Universe and proxy for metallicity in the extragalactic domain-- can be determined using optical CELs. To obtain precise values of the O abundance, we need to apply the so-called direct method \citep{Dinerstein:90}, which requires a determination of the electron temperature, \tel, through the measurement of faint auroral CELs of certain ions, on which the intensity depends exponentially on \tel. When the direct method cannot be applied, metallicity is then estimated through the so-called strong line methods based on the intensity of bright nebular CELs, which are  calibrated using empirical abundances \cite{Pagel:79, Pettini:04, Pilyugin:16} or photoionization models \cite{McGaugh:91,Kewley:08}. These are routinely used in surveys dedicated to studying large samples of faint SFDGs \cite{Tremonti:04, Izotov:06, Marino:13} or faint/high-metallicity {\hii} regions within spiral galaxies \cite{Sanchez:14, Belfiore:17,Groves:23}. It is important to note that, similar to local analogues,  
the abundances of primeval galaxies observed with JWST can now be  determined using the direct method  
\cite{ArellanoCordova:22,Schaerer:22,Curti:23,Trump:23,Sanders:24}. 

In addition to CELs, nebular spectra also feature recombination lines (RLs), whose intensity depends very weakly on {\tel}. The brightest RLs are those of \hi\ or \hei\, but the \oii, \oi, or \cii\ RLs are also present although very faint, at most 10$^{-3}$ to 10$^{-4}$ times the intensity of H$\beta$. Since 1942 \cite{Wyse:42}, we know that optical RLs provide systematically higher abundances than the CELs of the same ion. This is the so-called abundance discrepancy (AD), quantified by the AD factor (ADF, difference between the  abundances derived from RLs and CELs). In {\hii} regions, the ADF of O$^{2+}$, ADF(O$^{2+}$), is a factor between 2 and 5 \cite{GarciaRojas:07, ToribioSanCipriano:17}. Such a dramatic difference calls into question whether the routine methods based on CELs are providing the true metallicities of star-forming regions. The origin of the AD has been largely unknown, but \cite{MendezDelgado:23a}, found a correlation between the ADF and temperature fluctuations (variations in the spatial distribution of \tel, parameterized by \ts\cite{Peimbert:67}). \cite{MendezDelgado:23a} also demonstrated that the presence of \ts\ only affects in a significant way to the high ionization zone of  nebulae (where O$^{2+}$ ion is located) and therefore only the value of \tel([\oiii]) should be affected by \ts, not \tel([\nii]) or \tel([\oii]). In conclusion, the effect of \ts\ onto \tel\ determinations implies that abundances determined using \tel([\oiii]) --the only \tel\ determinations  available for SFDGs in the vast majority of cases-- could be severely underestimated, especially in low-metallicity {\hii} regions and SFDGs. The AD/\ts\ problem also affects the application of strong-line methods such as R23, N2 and O3N2, which are calibrated either on CEL-based \tel\ measurements or on photoionization models. They yield mutually inconsistent abundance scales and can differ by $\gtrsim$ 0.5 dex at fixed line ratios, propagating large systematics into the mass-metallicity and other fundamental metallicity relations \cite{Maiolino:19, Kewley:19ARAA}. For all the aforementioned reasons, it is essential to obtain more RL-based abundance determinations in SFDGs and EMPGs where available data are extremely limited~\cite{MendezDelgado:23a}.

 In addition to a dependence on \tel, abundance calculations also rely on precise determination of the electron density, \nel. In the last years, the use of 10m-class telescopes, has permitted the community to derive \nel\ from different CEL ratios ([\sii], [\oii], [\cliii], [\ariv], [\feiii]) in the spectra of extragalactic {\hii} regions and SFDGs, 
 moving beyond the typical determination solely based on [\sii] $\lambda$6717/$\lambda$6731. \cite{MendezDelgado:23b} 
 found that \nel\ tends to be underestimated when only 
 [\sii] diagnostic is used; this is caused by the non-linear \nel\ dependence of the different line ratios, suggesting the ubiquity of density inhomogeneities in {\hii} regions. The average underestimate in the available local sample is $\sim$300 cm$^{-3}$, which introduces systematic overestimates in the {\tel} calculations. In general, the \nel\ underestimate has a small impact on  abundances derived from optical CELs, being less than $\sim$0.1 dex. However, the \nel\ effects are critical when using infrared fine structure CELs. Although those puzzling results are currently restricted to local, low-\nel\ (10$^1$-10$^3$ cm$^{-3}$), and relatively bright objects, they could have a more significant effect when using the rest-frame UV-optical spectra of high-$z$ galaxies. Recent work suggests a cosmic evolution of \nel, reaching values on the order of 10$^4$-10$^5$ cm$^{-3}$ in high-$z$ galaxies \cite{Davies:21, Isobe:23}. In that case, the adopted value of the standard [\sii] diagnostic is largely inadequate and can result in incorrect abundance determinations biased toward lower values, thereby altering our understanding of the chemical evolution of primeval galaxies. A further need to study the \nel\ and \tel\ structure of local SFDGs in much greater detail arises from unexpected measurements of extremely high N/O ratios in high-$z$ galaxies \cite{Bunker:23,Cameron:23,Schaerer:24}. While the origin of N overabundance challenges our understanding of its nucleosynthesis in metal-poor systems, the impact of \nel\ on metallicity has important consequences for key scaling relations, such as the mass-metallicity relation that is closely linked to galaxy formation and evolution. 

Another frontier goal of observational astrophysics is to understand the epoch of reionization (EoR) and its connection to the young stellar populations of the first  galaxies. In the EoR, galaxies consistently exhibit stronger nebular emission lines due to their bursty star formation and possible black hole (BH) activity, and their ionized ISM show more extreme properties than found locally \cite{Stark:16}. 
They have larger EWs, \nel\ and \tel, reflecting hard radiation fields and more extreme ISM conditions, as evidenced by strong high ionization lines, e.g. {\heii}, [\nev] or {\civ}. 
The combination of these properties with a highly turbulent ISM due to bursty star formation and strong feedback, mostly from massive stars and SNe, facilitate the escape  of ionizing photons to the intergalactic medium, thus contributing to reionization. 

However, the origin of such properties and the unambiguous characterization of the ionizing sources leading to such strong ionization lines, remains unknown. In particular, the high-energy ionizing continuum ($>$54 eV) is completely unconstrained, especially in the most metal-deficient SFDGs, where the high-ionization phenomenon is expected to be common. This translates to an alarming issue when interpreting the spectra of high-z galaxies (e.g. \cite{Cleri:25}). Moreover, understanding the physics behind the nebular \heii\ $\lambda$4686 line is notably relevant as this line is considered a potential diagnostic of the elusive Pop III stars and their primeval hosts, which are in turn believed to have strongly contributed to the cosmic reionization \citep{Schaerer:08,Maiolino:24}.

\section{The Science Challenge}

The accumulated knowledge of local SFDGs has been largely built from samples accessible from the Northern Hemisphere, which allows observations of $\sim$70\% of the known Local Volume SFDGs (e.g. from \cite{Karachentsev2013}). Many classical SFDGs, including well-studied nearby analogs to high-$z$ galaxies, such as I\,Zw\,18, II\,Zw\,40, NGC\,2366/Mrk71, NGC\,1569, 
Leo\,P, CGCG\,007-025, and SBS0335$-$052E, have been the subject of deep observational campaigns from northern telescopes, consolidating their role as reference objects. Furthermore, the main surveys that identified and classified most known SFDGs primarily covered the Northern Hemisphere, including the historical Palomar Observatory Sky Survey (POSS) \cite{Sargent:70}, the  Sloan Digital Sky Survey (SDSS) \cite{Izotov:06}, or the most recent DESI and J-PAS spectrophotometric surveys \cite{Zinchenko2024, Gimenez-Alcazar2025}, which can provide the most complete local samples. Therefore, the density of high-quality targets, systematic catalogs, and historical continuity of observations has favored a better understanding of northern SFDGs, including the best analogues to high-z galaxies. However, nearby iconic prototypes of extreme SFDGs, such as I\,Zw\,18, Leo P or SBS0335-052E, 
widely studied with exquisite datasets from the ground and from space, and well visible from the North, still lack optical spectroscopic observations with sufficient depth to measure RLs of heavy elements (multiplet 1 of \oii\ at about $\lambda$4650 and \cii\ $\lambda$4267) or the faintest CELs (such as [\nii] $\lambda$5755, [\cliii] $\lambda\lambda$5518,5538, [\feiii] $\lambda$4702), all of which are needed to perform an adequate exploration of the presence and effects of \tel\ and \nel\ inhomogeneities and their impact on chemical abundances. 

Similarly, the two nearby starbursting dwarfs I~Zw\,18 and SBS\,0335$-$052E  are the most metal-poor (few \% solar) {\heii} and  [\nev] emitters at $z=0$  \cite{Kehrig:11}. However, unveiling the origin of such intrinsically faint high-ionization lines  remains a strong challenge and, despite many observational and theoretical attempts, the source of the high-ionizing SED (E > 54 eV) keeps challenging current stellar models in these galaxies (e.g., \cite{Eldridge:22}). The possibility remains that nebular high-ionizing lines (e.g. \heii, [\nev]) are powered by predicted peculiar hot massive stars (e.g. Pop III-like, chemically homogeneous evolving stars, see e.g. \cite{Kehrig:15,Kehrig:18,Kehrig:21,Arroyo-Polonio:25,Szecsi:25}). In I\,Zw\,18, direct measurements of $T_e$[\oiii] values of $\sim$22000~K are spatially coincident with the \heii\ ionizing gas \cite{Kehrig:16}, and discrepancies between observed and theoretical O$^{3+}$/H$^+$ values \cite{Rickards:25} indicate the existence of additional ionizing sources other than the conventional ones (e.g., WRs, X-ray sources).

A tenfold increase in the photon-collecting capacity of a 30m-class telescope --coupled with a high-throughput optical spectrographs and AO-fed IFUs-- in La Palma could undoubtedly provide detections and reliable measurements of the aforementioned relevant but faint RLs and CELs in sizeable samples of local SFDGs, EMPGs, and {\hii} regions, overcoming the present limitation to a handful of bright objects. This will allow us to quantify \nel\ and \tel\ inhomogene\textbf{}ities and the ADF, shown to be crucial for avoiding severe underestimates of the O abundance especially at low metallicity, and to derive an empirical RL-based calibration of classical strong-line indicators (e.g. O23, O3N2, N2) and photoionization models across a wide range of ionization parameter and excitation.  These deeper observations would yield unprecedented constraints on the nature, ionization and true chemical abundances of SFDGs, and are necessary to recalibrate strong line methods. Recalibration of the rest-frame optical/UV diagnostics used for galaxies at $z \gtrsim 2-10$ is essential to provide a robust metallicity baseline and interpret JWST and ELT surveys, and to reassess the cosmic evolution of the mass–metallicity relation and the baryon cycle. 



\begin{multicols}{2}
\renewcommand{\bibfont}{\footnotesize}  

\end{multicols}

\end{document}